\documentclass[a4paper]{jpconf}
\usepackage{graphicx}
\begin{document}
\title{Current status of the CLIO project}

\author{K Yamamoto$^1$,
T Uchiyama$^1$, S Miyoki$^1$, M Ohashi$^1$, K Kuroda$^1$,\\ 
H Ishitsuka$^1$, T Akutsu$^1$, 
S Telada$^2$, T Tomaru$^3$, T Suzuki$^3$, N Sato$^3$, Y Saito$^3$, 
Y Higashi$^3$,
T Haruyama$^3$,
A Yamamoto$^3$, T Shintomi$^4$, \\
D Tatsumi$^5$, M Ando$^6$, 
H Tagoshi$^7$, N Kanda$^8$, N Awaya$^8$, \\
S Yamagishi$^8$, H Takahashi$^9$,  
A Araya$^{10}$, A Takamori$^{10}$, S Takemoto$^{11}$, T Higashi$^{11}$, 
H Hayakawa$^{11}$, 
W Morii$^{12}$ and J Akamatsu$^{12}$}

\address{$^1$ Institute for Cosmic Ray Research, The University of Tokyo, 
Kashiwa, Chiba 277-8582, Japan}
\address{$^2$ National Institute for Advanced Industrial Science 
and Technology, Tsukuba, 
Ibaraki 305-8563, Japan}
\address{$^3$ High Energy Accelerator Research Organization, Tsukuba, 
Ibaraki 305-0801, Japan}
\address{$^4$ Advanced Research Institute for the Sciences and Humanities, 
Nihon University, 
Chiyoda-ku, Tokyo 102-0073, Japan}
\address{$^5$ National Astronomical Observatory of Japan, Mitaka, 
Tokyo 181-8588, Japan}
\address{$^6$ Department of Physics, The University of Tokyo, Bunkyo-ku, 
Tokyo 113-0033, Japan}
\address{$^7$ Department of Earth and Space Science, 
Graduate School of Science, Osaka University, 
Toyonaka, Osaka 560-0043, Japan}
\address{$^8$ Department of Physics, Graduate School of Science, 
Osaka City University, Sumiyoshi-ku, Osaka, 
Osaka 558-8585, Japan}
\address{$^9$ Department of Management and Information Systems Science,
Nagaoka University of Technology, Nagaoka, Niigata 940-2188, Japan}
\address{$^{10}$ Earthquake Research Institute, The University of Tokyo, 
Bunkyo-ku, Tokyo 113-0032, Japan}
\address{$^{11}$ Department of Geophysics, Kyoto University, Sakyo-ku, Kyoto,  
Kyoto 606-8502, Japan}
\address{$^{12}$ Disaster Prevention Research Institute, Kyoto University, 
Uji, Kyoto 611-0011, Japan}

\ead{yamak@icrr.u-tokyo.ac.jp}

\begin{abstract}
CLIO (Cryogenic Laser Interferometer Observatory) is 
a Japanese gravitational wave detector project. 
One of the main purposes of CLIO is to demonstrate 
thermal-noise suppression by cooling mirrors 
for a future Japanese project, LCGT (Large-scale Cryogenic Gravitational 
Telescope). 
The CLIO site is in Kamioka mine, as is LCGT. 
The progress of CLIO between 2005 and 2007 
(room- and cryogenic-temperature experiments) is introduced 
in this article. 
In a room-temperature experiment, 
we made efforts to improve the sensitivity.
The current best sensitivity at 300 K is about 
$6 \times 10^{-21} /\sqrt{\rm Hz}$ around 400 Hz. 
Below 20 Hz, the strain (not displacement) sensitivity is comparable 
to that of LIGO, 
although the baselines of CLIO are 40-times shorter 
(CLIO: 100m, LIGO: 4km). 
This is because seismic noise 
is extremely small in Kamioka mine. We operated the interferometer 
at room temperature for 
gravitational wave observations. We obtained 86 hours of data. 
In the cryogenic experiment, 
it was confirmed that 
the mirrors were sufficiently cooled (14 K). 
However, we found that the radiation shield ducts transferred 300K 
radiation into the cryostat more effectively than we had expected. 
We observed that noise caused by pure aluminum wires 
to suspend a mirror was suppressed 
by cooling the mirror.    
\end{abstract}

\section{Introduction}

Observations using several interferometric gravitational wave detectors 
(LIGO \cite{LIGO}, 
Virgo \cite{VIRGO}, GEO \cite{GEO}, TAMA \cite{TAMA}) 
on the ground are presently in progress. 
It is possible for the current LIGO interferometers to detect a chirp wave 
of a neutron-star 
binary coalescence at about 15 Mpc far from Earth \cite{LIGOcurrent}.
Since  
this observational distance must be extended to 200 Mpc in order to detect 
a few chirp events every year,  
some future projects have been proposed: 
Advanced LIGO (U.S.A.) \cite{advancedLIGO} and LCGT (Japan) \cite{LCGT}.    
The LCGT (Large-scale Cryogenic Gravitational Telescope) project  
has some features that the current and other future projects do not have. 
One of these is 
cryogenic mirrors (below 20 K) and suspensions in order 
to reduce thermal noise.
Another feature is that the site is 1000 m underground (Kamioka mine) 
because of small seismic motion. As a prototype of LCGT, 
the CLIO (Cryogenic Laser Interferometer Observatory) project 
\cite{CLIO2002,CLIOAmaldi5,CLIOAmaldi6} is now in progress. 
The goal of CLIO is to  
construct an interferometer in Kamioka mine and to demonstrate 
thermal-noise suppression by cooling the mirrors. 
The CLIO and LCGT projects include the construction 
and operation of apparatus for geophysics 
\cite{CLIO2002,CLIOAmaldi5,CLIOAmaldi6,Takemoto1,Takemoto2}.   

\begin{figure}[h]
\includegraphics[width=18pc]{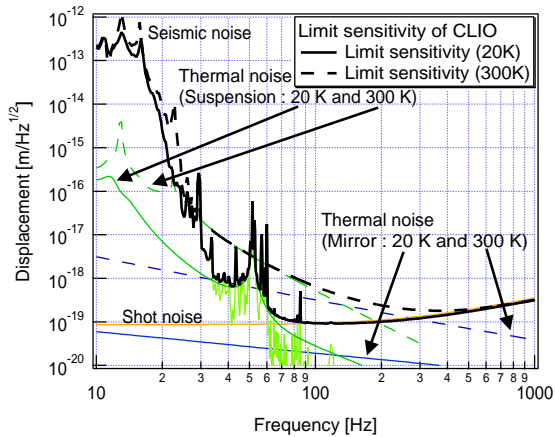}\hspace{2pc}%
\begin{minipage}[b]{18pc}\caption{\label{CLIOlimit}
Limit sensitivity of the CLIO interferometer. 
This limit consists of the fundamental noise of the interferometers: 
shot noise, seismic noise, thermal noise of the suspensions and mirrors. 
The thick black dashed and the solid 
lines are the limit sensitivity at room and cryogenic
temperatures, respectively. The limit sensitivity between 30 Hz and 400 Hz
becomes about 10-times better when the mirrors and suspensions are cooled.}
\end{minipage}
\end{figure}

Details of the CLIO interferometer are given 
in Refs. \cite{CLIO2002,CLIOAmaldi5,CLIOAmaldi6}.
An outline is introduced here. The site, Kamioka mine 
(Hida city, Gifu prefecture, Japan)  
is 220 km west of Tokyo (TAMA site). In this mine, 
there is the world-famous water 
Cherenkov neutrino detector, Super-Kamiokande \cite{SK}. 
The vertical distance from the level of Super-Kamiokande
and CLIO to the top of this mountain is about 1000 m. 
The horizontal distance from a mine entrance to the 
CLIO site is about 2000 m. Seismic vibration in Kamioka mine 
is about 100-times less than 
that in the suburbs of Tokyo \cite{Araya,YamamotoAmaldi6}. 
At this silent site, stable operation with lower 
seismic noise is possible \cite{Sato}. 
The CLIO interferometer has 
two 100 m length Fabry-Perot cavities 
(LCGT: 3 km) which consist of 
four sapphire mirrors. These mirrors 
and their suspensions are cooled by silent pulse-tube cryocoolers 
developed for CLIO 
\cite{YamamotoAmaldi6,Cryocooler1,Cryocooler2,Cryocooler3}. 
The temperature of these mirrors must be below 20 K. 
Light from a laser source passes through a 10 m mode cleaner, 
and is divided 
by a beam splitter for the 
two cavities. A power-recycling mirror does not exist. 
The power on the beam splitter is 0.5 W. 
Reflected beams by the two cavities are not recombined. 
The inline cavity, into which the light 
transmitted by the beam splitter goes, is used for laser frequency 
stabilization. 
The length of the 
other cavity, the perpendicular one, is controlled 
so as to keep the light 
in this cavity resonant. 
Feedback of this control system includes gravitational wave signals 
\cite{lockedFP}. 
Figure \ref{CLIOlimit} shows the limit sensitivity of the CLIO interferometer. 
This limit consists of the fundamental noise of the interferometers: 
shot noise, seismic noise, thermal noise of the suspensions and mirrors. 
The thick black dashed and the solid 
lines are the limit sensitivity at room and cryogenic temperatures, 
respectively. 
At room temperature, the limit sensitivity below 30 Hz and above 400 Hz 
is dominated by the seismic and shot noise, respectively. 
Between 30 Hz and 400 Hz, the sensitivity is limited by the thermal noise.
The limit sensitivity in this frequency region becomes about 10-times better 
when the mirrors and suspensions are cooled. 

The CLIO project began in 2002. In autumn of this year, tunnels 
for CLIO were completed. 
Just before the previous Amaldi conference (June 2005), 
all vacuum chambers, pipes, pumps, cryostat, and cryocoolers were installed 
and assembled 
\cite{CLIOAmaldi6,UchiyamaAmaldi6}. 
After the previous Amaldi conference, optical and suspension systems 
were installed. 
Moreover, we made efforts for room- and cryogenic-temperature experiments.
In the room-temperature experiment, we improved the sensitivity and 
operated the interferometer for gravitational wave observation. 
In the cryogenic experiment, 
we checked 
whether the cryostat systems cool the mirrors sufficiently, and found 
that cooling a mirror reduced the noise. 
In this article, the progress of the CLIO project 
after the previous Amaldi conference is introduced.  

\section{Room-temperature experiment}

\subsection{History of the room-temperature experiment}

After June 2005, we installed optics and suspensions. 
A one-arm cavity experiment began on September 2005. 
On 18 February 2006, the CLIO interferometer was fully locked. 
After July 2006, 
we made many efforts to improve the sensitivity. On 21 and 22 November, 2006, 
the data-acquisition systems were tested. 
On 13 December 2006, we obtained the current best sensitivity of CLIO. 
On February 2007, we operated the CLIO interferometer for one week 
to conduct gravitational wave observations.  
Here, the current best sensitivity and observations are discussed. 

\subsection{Current best sensitivity}

\begin{figure}[h]
\begin{minipage}{18pc}
\includegraphics[width=18pc]{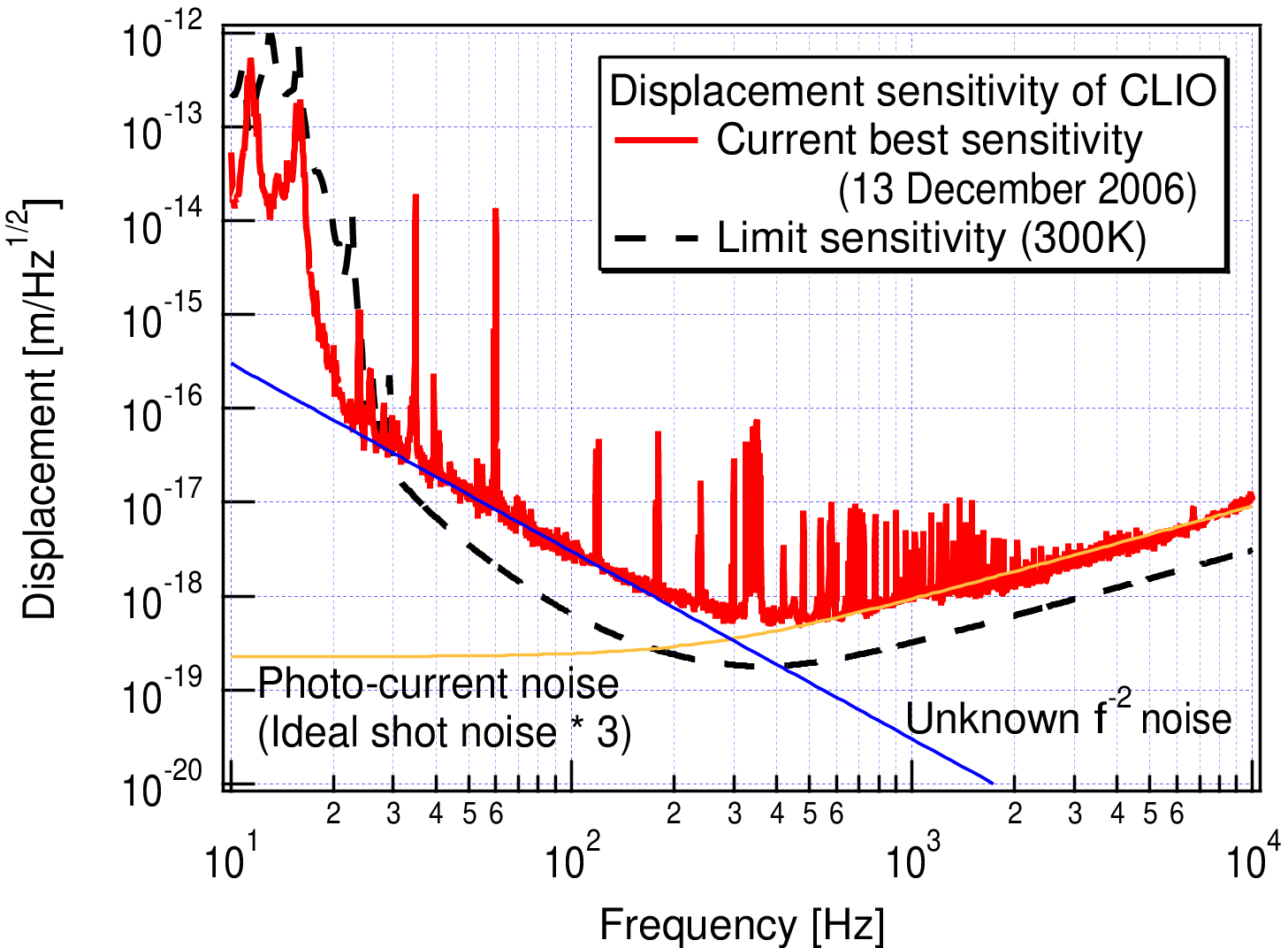}
\caption{\label{CLIOsens1}Current best displacement sensitivity of the 
CLIO interferometer at room temperature. 
The thick solid (red in online) and dashed (black in online) lines are 
the current best and limit sensitivity, respectively. 
Above 300 Hz, photo-current noise limits the sensitivity. 
This noise is 3-times larger than the ideal shot noise. 
Between 20 Hz and 300 Hz, unknown noise limits the sensitivity. 
Below 20 Hz, the sensitivity is limited by the seismic noise.}
\end{minipage}\hspace{2pc}%
\begin{minipage}{18pc}
\includegraphics[width=18pc]{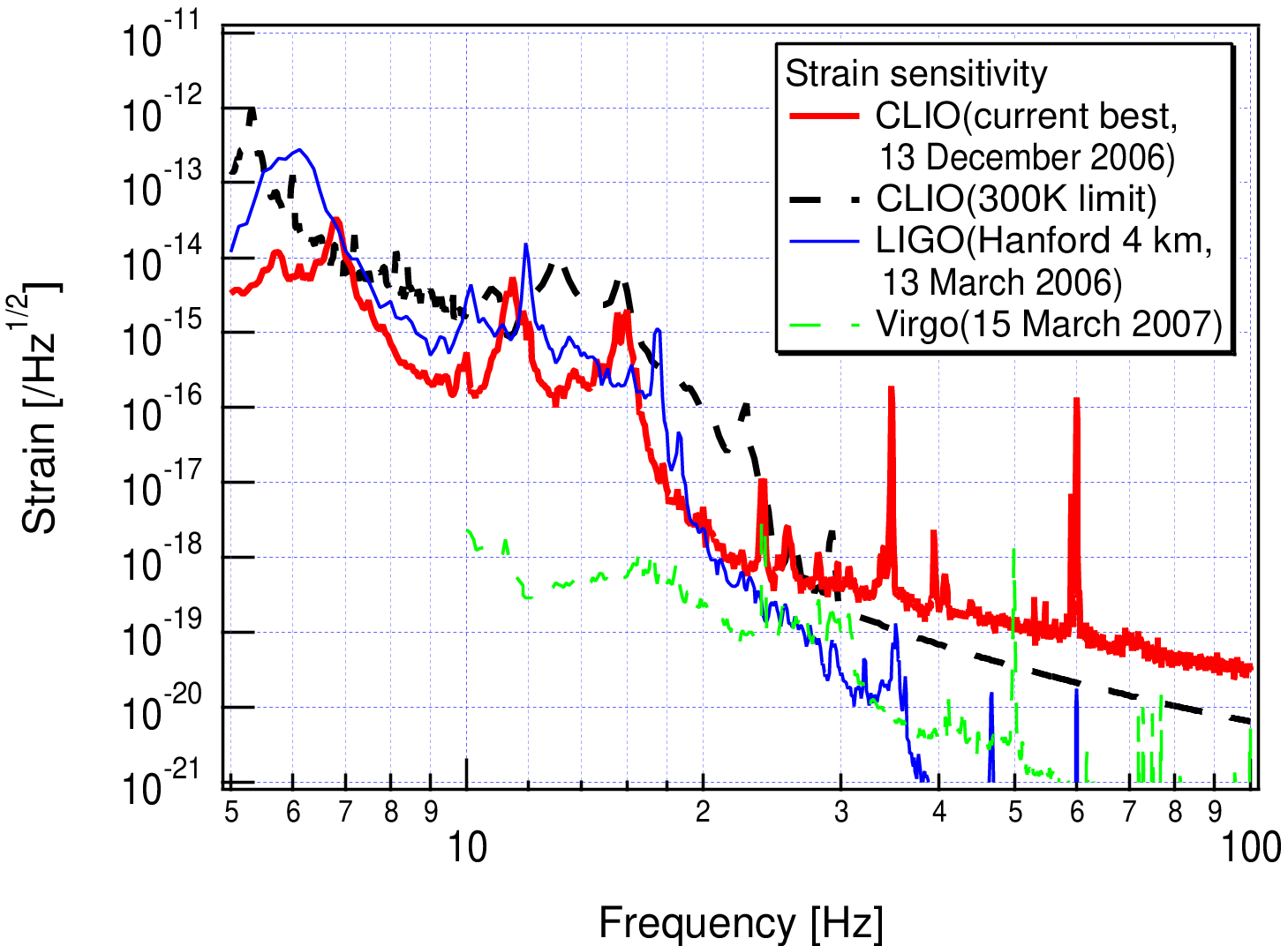}
\caption{\label{CLIOsens2}Current best strain sensitivity of 
the CLIO interferometer 
in a lower frequency region at room temperature.
The thick solid (red in online) and dashed (black in online) lines are 
the current best and limit sensitivity of CLIO, respectively. 
The thin solid (blue in online) and dashed 
(green in online) lines are the 
sensitivity of LIGO \cite{LIGOcurrent} 
and Virgo \cite{VIRGOcurrent}.}
\end{minipage} 
\end{figure}

Figure \ref{CLIOsens1} shows the current best displacement sensitivity of 
the CLIO interferometer. 
The thick solid (red in online) and dashed (black in online) lines are 
the current best and limit sensitivity, respectively. 
The strain best sensitivity 
around 400 Hz is about $6 \times 10^{-21} /\sqrt{\rm Hz}$. 
Above 300 Hz, photo-current noise limits the sensitivity. 
This noise is 3-times larger than the ideal shot noise.
Studies after this conference revealed that this was shot noise. 
The reason why the actual shot noise is different from the ideal one 
is now being investigated. 
Between 20 Hz and 300 Hz, 
the sensitivity is inverse proportional to the square of the frequency. 
This noise is unknown and several-times larger than the goal 
sensitivity. 
Below 20 Hz, the sensitivity is limited 
by the seismic noise. There is no alignment control noise, 
because alignment control systems had not been 
installed at that time. 
Figure \ref{CLIOsens2} shows the best strain sensitivity 
of the CLIO interferometer in a low-frequency region. 
Below 20 Hz, the strain (not displacement) sensitivity of CLIO 
is comparable to that of LIGO 
(thin solid line, blue in online) \cite{LIGOcurrent}; 
nevertheless, the arm length of CLIO is 40-times shorter 
(LIGO: 4 km). 
The Virgo strain sensitivity 
(thin dashed line, green in online) \cite{VIRGOcurrent}
is better than that of CLIO. 
However, the displacement sensitivity of CLIO is better 
between 20 Hz and 30 Hz (Virgo arm length: 3 km), 
although CLIO does not adopt 
low-frequency vibration-isolation systems,
as do Super Attenuators of Virgo. 
Such good sensitivity of the CLIO interferometer in the lower frequency region 
is because of extremely small seismic vibrations in Kamioka mine. 

\subsection{Observation}

\begin{figure}[h]
\includegraphics[width=18pc]{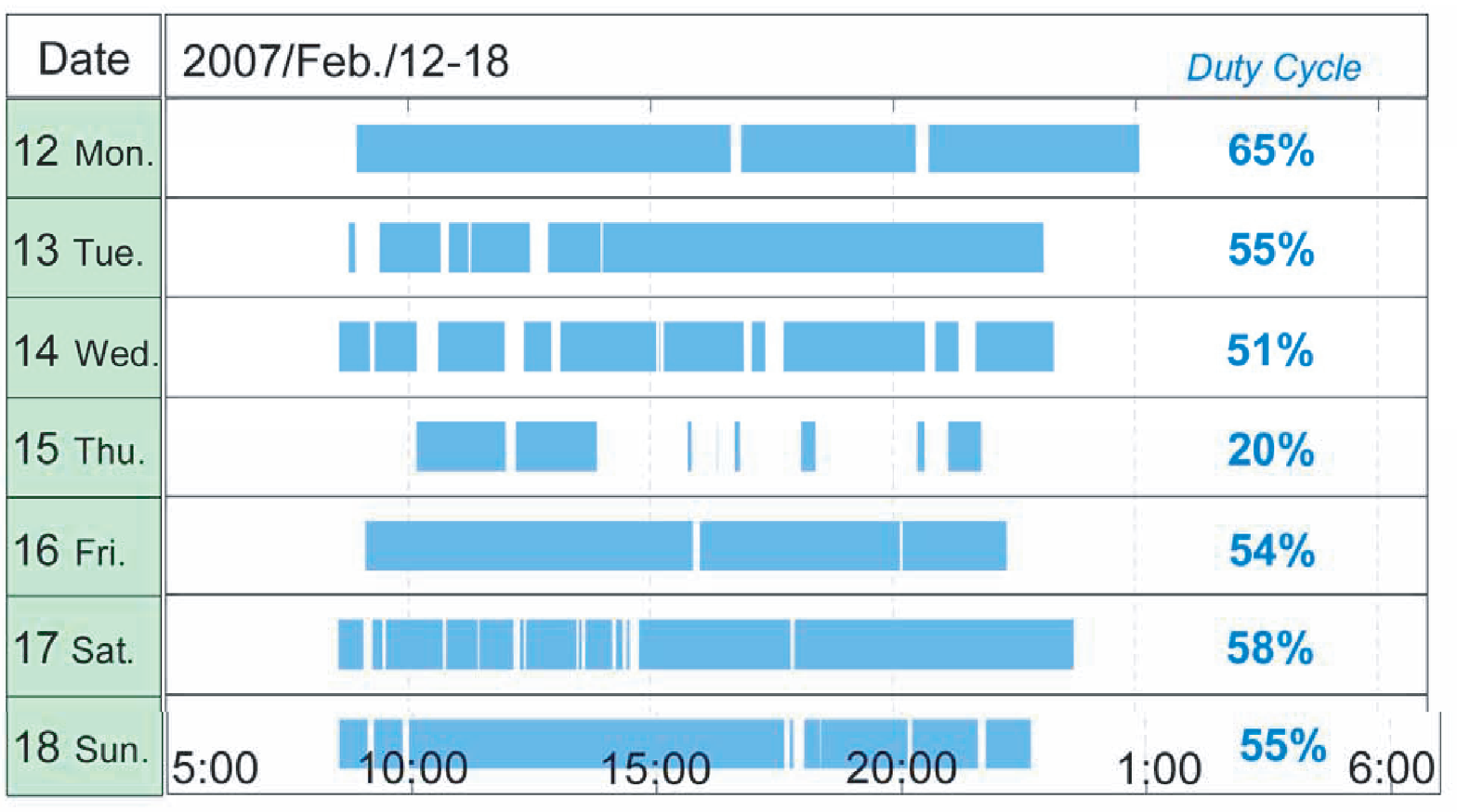}\hspace{2pc}%
\begin{minipage}[b]{18pc}\caption{\label{obstable}
CLIO interferometer status during the observation on February 2007. 
The horizontal axis shows time from 
5 am on one day to 6 am on the next day. The blue parts imply that 
the interferometer was locked. On the right-hand side of this table, 
the duty cycle on each day is shown. 
The total duty cycle was 51\%. 
The longest lock was about 9 hours (in the afternoon on 13 February).}
\end{minipage}
\end{figure}
We operated the CLIO interferometer at room temperature 
for observations. 
The data obtained at 300 K will be compared 
with those at the cryogenic temperature. 
The observation term was between 12 and 18 of February, 2007. 
The sensitivity in the observation was almost the same as the current best one.
The observable inspiral range of neutron-star binaries was 49.5 kpc 
(in a case of a optimum direction. 
The threshold of the signal-to-noise ratio was 10).
The storage data length was 86 hours. 
Data analysis is now in progress. Figure \ref{obstable} shows 
the status of the interferometer 
during the observation. The horizontal axis shows time.  
The blue parts imply that the interferometer was locked. 
The longest lock was about 9 hours (in the afternoon on 13 February).  
The duty cycle on each day is shown in fig. \ref{obstable}. 
The total duty cycle was 51\%. This duty cycle was not a disappointing result, 
because auto lock and alignment control systems were not installed. 

Any improvement for stable operation is necessary. 
For example, a lock of the interferometer was broken 
before 2 am every night, and never recovered. From 10 pm to 8 am, 
nobody was near the interferometer 
(at least two operators were near the interferometer 
from 8 am to 10 pm everyday). 
An auto lock systems must be installed for operation without operators. 
On 15 February, it was difficult to acquire a lock 
because a terrible storm occurred. 
Strong wind increased the seismic noise between 0.1 Hz and 1 Hz. 
The duty cycle on this day was only 20\%. 
For lock acquisition and stable locks on such a terrible storm day, 
systems to tune the coil-magnet actuator efficiency 
and to control the upper mass (not the mirror) are necessary. 
Alignment control systems for slow drift 
cancellation to keep the best sensitivity must also be considered. 

\section{Cryogenic-temperature experiment}

\subsection{Step of cryogenic-temperature experiment}

In the room-temperature experiment, the mirrors were suspended 
by 400 mm length Bolfur \cite{Bolfur} wires. 
This material has a high tensile strength. 
Bolfur wire is not useful in  
cryogenic experiments because of low thermal conductivity. 
In the final step of the cryogenic experiment, sapphire fibers will 
be adopted as LCGT suspensions \cite{LCGT}.
A sapphire fiber has high thermal conductivity \cite{Tomarusapphire} 
and small mechanical loss  
\cite{Uchiyamasapphire}. However, it is expensive and fragile.
Thus, in the first step 
of the cryogenic experiment, pure aluminum wires that are inexpensive and 
not fragile are used \cite{UchiyamaAmaldi6}. 
A pure aluminum fiber has a high thermal conductivity. 
A fault is large mechanical loss.  
We are now in the first step of the cryogenic experiment. 
Two interesting topics found 
in this first-step experiment are introduced: a mirror cooling test 
and a sensitivity improvement.   

\subsection{Mirror cooling test}

We operated the cryocoolers in order to confirm that the mirrors 
could be cooled sufficiently \cite{UchiyamaAmaldi6}. 
The results are 
summarized in Table \ref{coolingtest} (the perpendicular front mirror 
was cooled after this Amaldi conference). 
All of the mirror temperatures became below 14 K 
within about one week (168 hours).
Since the mirror temperature must be less than 20 K, 
the cryogenic system worked well 
(other parts of the cryostat were also sufficiently cooled). 
\begin{table}[h]
\caption{\label{coolingtest}Result of the mirror cooling test.}
\begin{center}
\begin{tabular}{lccc}
\br
 &Cooling time&Mirror temperature&Heat into mirror\\
\mr
Inline front mirror&174 hours&13.4 K&36 mW\\
Inline end mirror&176 hours&13.5 K&40 mW\\
Perpendicular front mirror&193 hours&13.8 K&29 mW\\
Perpendicular end mirror&164 hours&12.5 K&62 mW\\
\br
\end{tabular}
\end{center}
\end{table}

\begin{figure}[h]
\begin{minipage}{18pc}
\includegraphics[width=18pc]{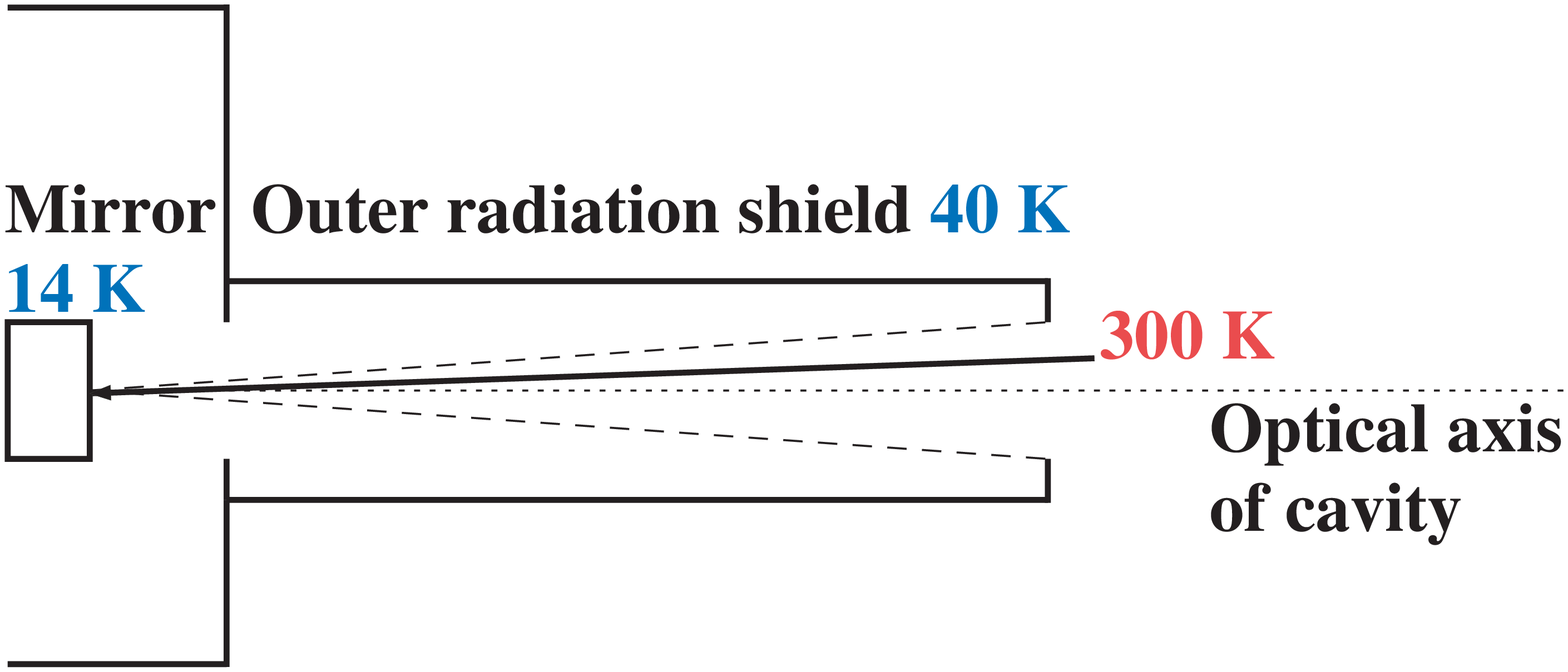}
\caption{\label{duct1}Model used in a previous estimation of 
heat into a mirror. 
The mirror and a part of the optical path are surrounded 
by a radiation shield to prevent 
300K radiation from attacking the mirror. 
We considered only 300K radiation entering the mirror directly.}
\end{minipage}\hspace{2pc}%
\begin{minipage}{18pc}
\includegraphics[width=18pc]{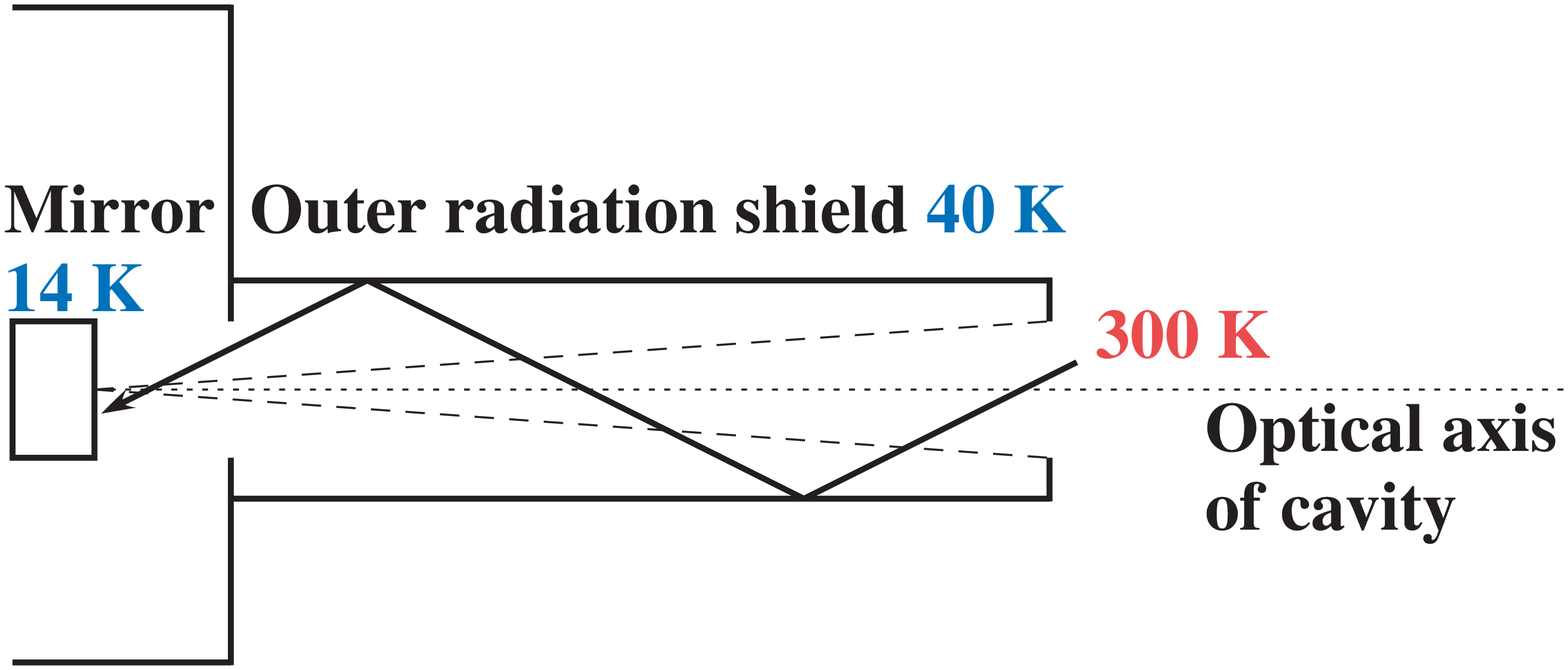}
\caption{\label{duct2}Revised model for estimating the heat 
into the mirror. 
The shield does not absorb, but reflects radiation, 
because shields are not black bodies.
300K radiation reflected by the 
duct goes into the mirror. This radiation power is 1000-times larger 
than that of the direct radiation shown in fig. \ref{duct1}, 
although this contribution was neglected in a previous estimation
\cite{TomaruAmaldi7}.}
\end{minipage} 
\end{figure}
Although the mirrors were cooled sufficiently, we found a problem of 
heat into the mirrors. 
The heat evaluated from this experiment was about 1000-times larger 
than the expected value (the heat  
in Table \ref{coolingtest} was obtained after extra baffles 
and radiation shields were installed in the cryostat).
When the expected 
value was calculated, we considered the model shown in fig. \ref{duct1}. 
A mirror and a part of 
the optical path are surrounded by a radiation shield to prevent 
300K radiation from attacking the mirror. 
We considered only 300K radiation entering the mirror directly, 
as shown in fig. \ref{duct1}. 
However, our recent investigation \cite{TomaruAmaldi7} revealed that 
the shield does not absorb, but reflects radiation 
(shields are not black bodies). 
300K radiation reflected by the 
duct goes into the mirror (fig. \ref{duct2}). 
The contribution of 
this reflected radiation is large, although we have neglected it. 
We must consider this problem in the design of the cryostat for LCGT. 
Details of this problem and a solution for LCGT 
are introduced in Ref. \cite{TomaruAmaldi7}.

\subsection{Sensitivity of cryogenic interferometer}

\begin{figure}[h]
\begin{minipage}{18pc}
\includegraphics[width=18pc]{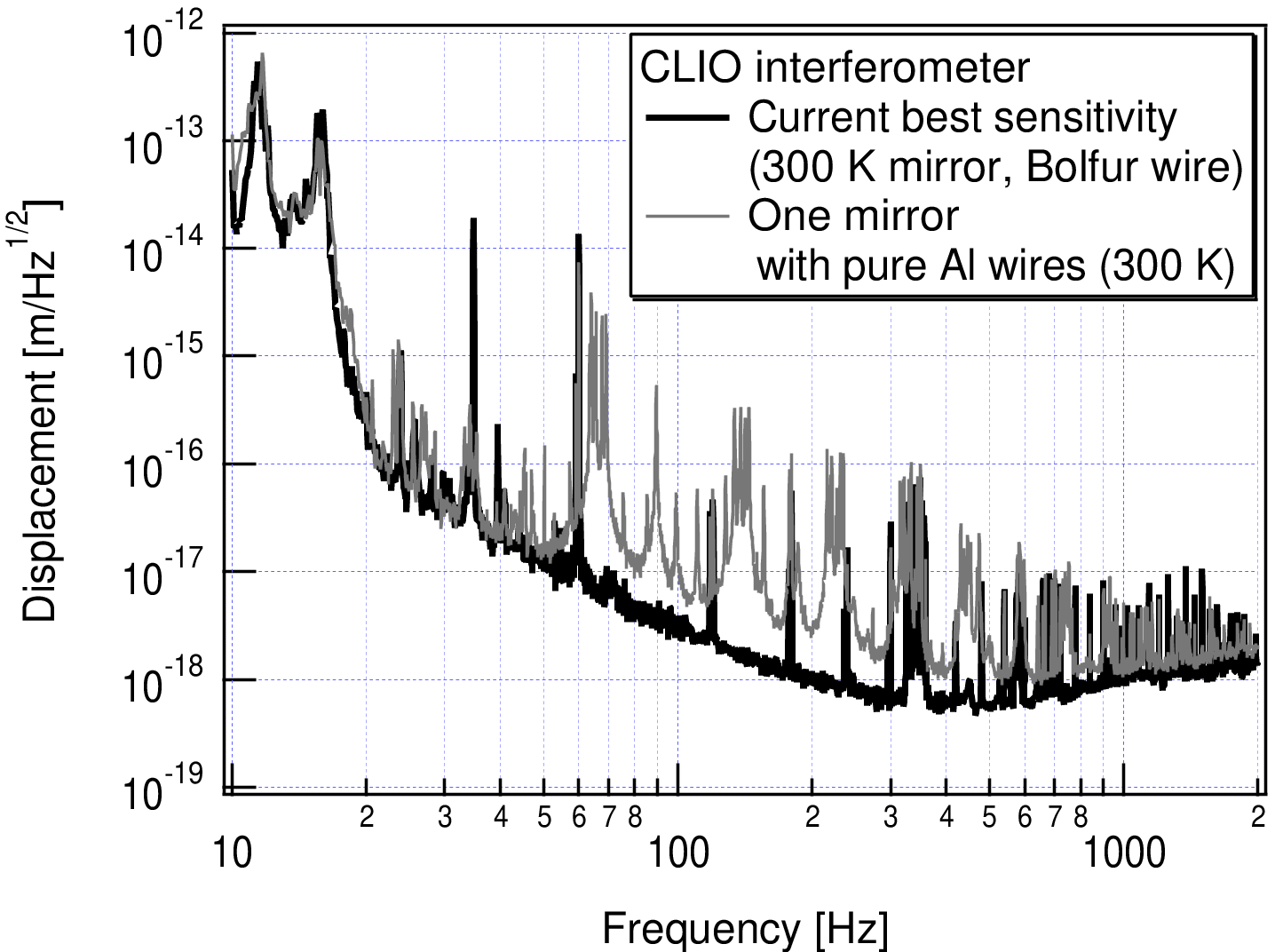}
\caption{\label{cryosens1}Sensitivity at room temperature. 
The thick black 
and thin grey lines are the current best sensitivity 
with the Bolfur wires at 300 K and the room-temperature sensitivity 
when a mirror was suspended by pure aluminum wires (1 mm in diameter).}
\end{minipage}\hspace{2pc}%
\begin{minipage}{18pc}
\includegraphics[width=18pc]{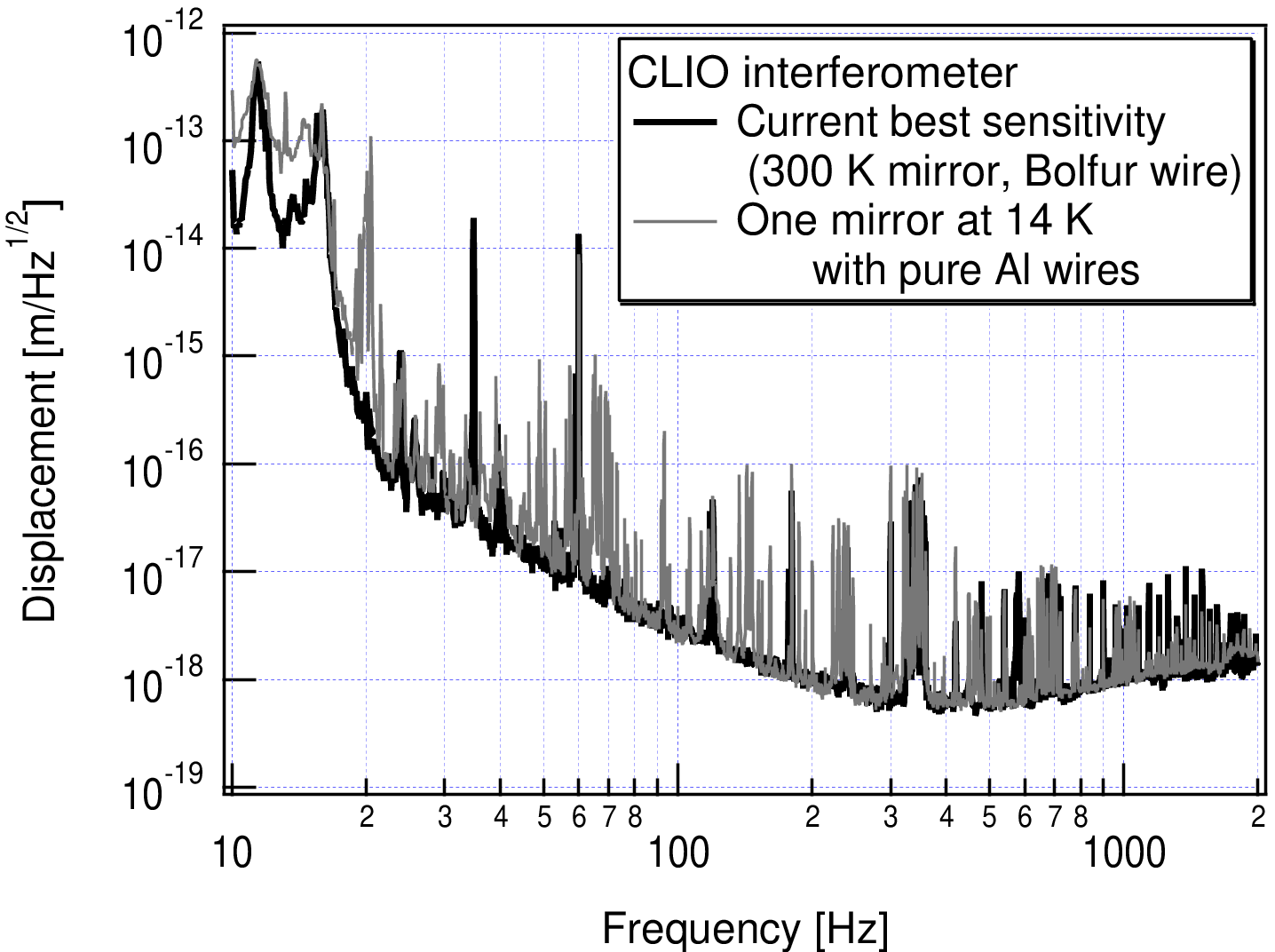}
\caption{\label{cryosens2}Sensitivity at room and 
cryogenic temperatures. 
The thick black and thin grey lines are the current best sensitivity 
with the Bolfur 
wires at 300 K and the sensitivity 
when a mirror was suspended by aluminum wires 
(1 mm in diameter) and cooled (14 K).}
\end{minipage} 
\end{figure}

In order to investigate the sensitivity at cryogenic temperature, 
we replaced the Bolfur wires by pure 
aluminum wires (1 mm in diameter) in a suspension. 
Only this suspension was cooled.
We measured the sensitivity of the interferometer. 
The results are summarized in figs. \ref{cryosens1} and \ref{cryosens2}. 
The thick black line in figs. \ref{cryosens1} and \ref{cryosens2} 
is the current best sensitivity with the Bolfur wires at 300 K.
The thin grey line in fig. \ref{cryosens1} 
is the room-temperature sensitivity with the pure aluminum wires. 
We found that the noise between 50 Hz and 800 Hz increased. 
This noise was caused by the pure aluminum wires. The thin grey line 
in fig. \ref{cryosens2} is the sensitivity 
when the mirror temperature was 14 K. 
The noise caused by the aluminum wires 
between 50 Hz and 800 Hz was suppressed by cooling this mirror. 
It must be noted that the floor level at 14 K 
was comparable to the best sensitivity at 300 K, 
although the cooled mirror and suspension had heat links 
\cite{UchiyamaAmaldi6}
and lines for power supplies and signal probes of thermometers. 

We found that the noise produced by the pure aluminum wires 
decreased when the 
mirror and suspension were cooled. In order to answer whether we observed 
the suspension thermal 
noise and the reduction by cooling, an investigation is necessary. 
This is because we do not understand why many peaks appear 
in the spectrum density in figs. \ref{cryosens1} and \ref{cryosens2}, 
i.e. what resonant modes they represent. 

\begin{figure}[h]
\includegraphics[width=18pc]{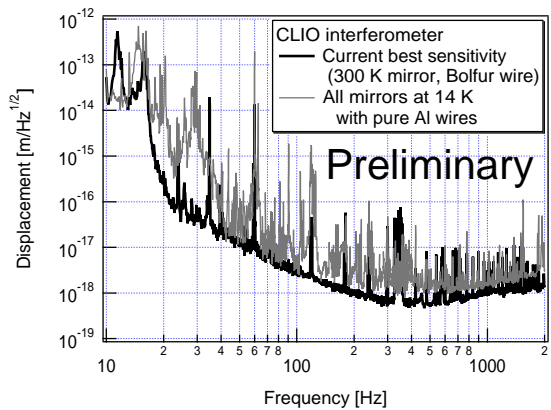}\hspace{2pc}%
\begin{minipage}[b]{18pc}\caption{\label{CLIOallcool}
(Preliminary) sensitivity with all four cooled mirrors. 
The thick black and thin grey lines are the current best sensitivity 
with the Bolfur wires at 300 K and the sensitivity 
when all of the mirrors were cooled 
with pure aluminum wires (0.5 mm in diameter), 
respectively.}
\end{minipage}
\end{figure}

After this conference, all of the sapphire mirrors were suspended 
by 0.5 mm diameter pure aluminum wires and cooled. 
The (preliminary) sensitivity with the four cooled mirrors 
is shown in fig. \ref{CLIOallcool} (thin grey line).
This is the first obtained sensitivity 
of a fully cooled interferometric gravitational 
wave detector. 
Although there were many peaks, 
the floor level between 40 Hz and 100 Hz was comparable to 
the current best sensitivity at room temperature (thick black line).
The current best sensitivity at 300 K is not limited 
by the thermal noise in this 
frequency region, probably. 

\section{Future work}

There are two kinds of future work: sensitivity improvement 
and long-term operation at 
cryogenic temperature. Regarding the former, we will try to reduce 
the shot noise   
(above 300Hz) and to investigate the unknown noise 
(between 20 Hz and 300 Hz). 
In the latter, we must install some apparatus for long observations: 
auto lock, 
alignment control, drift cancel, digital control systems, etc.   

\section{Summary}

The CLIO (Cryogenic Laser Interferometer Observatory) project 
\cite{CLIO2002,CLIOAmaldi5,CLIOAmaldi6} is now in progress. 
One of main purposes of CLIO is to 
demonstrate thermal-noise suppression by cooling mirrors 
for the future Japanese 
project, LCGT \cite{LCGT}. The CLIO and LCGT sites are 
in Kamioka mine, where  
seismic noise is extremely small. After the previous Amaldi conference 
(recent two years), 
we made efforts in room- and cryogenic-temperature experiments.

In the room-temperature experiment, we improved the sensitivity and 
operated the interferometer for gravitational wave observations. 
At the end of 2006, we obtained the current best sensitivity.  
The strain sensitivity 
around 400 Hz is about $6 \times 10^{-21} /\sqrt{\rm Hz}$. 
Above 20 Hz, the current best sensitivity is a few or several-times larger 
than the limit sensitivity at room temperature.
Since the seismic vibrations are small in Kamioka mine, 
the strain (not displacement) sensitivity below 20 Hz
is comparable to that of LIGO, although the baseline length of CLIO 
is 40-times shorter.
In February 2007, we operated the CLIO interferometer 
at room temperature for observations. 
We obtained 86 hours of data. These data are now being analyzed. 

In the cryogenic experiment, we confirmed that the mirror temperatures 
became about 14 K, which was 
lower than the goal temperature of 20 K. However, 300K radiation 
invading through the shield ducts is larger than 
that in the original design. This problem
and a solution for LCGT are introduced in Ref. \cite{TomaruAmaldi7}. 
We found that noise caused by aluminum wires used to suspend a mirror 
could be suppressed by 
cooling this mirror. An investigation is necessary to check 
whether we observed the suspension thermal noise and the suppression 
by cooling. 
  
The main topics of future work are sensitivity improvement and 
long-term observation at cryogenic temperature. 

\ack

This project was supported in part by Grant-in-Aid for Scientific Research 
on Priority Areas of the Ministry of Education, Culture, Sports, Science 
and Technology 
(13048101) and is supported in part 
by Grant-in-Aid for Scientific Research (A) of 
Japan Society for the Promotion of Science (18204021).

\end{document}